\documentclass[aps,twocolumn]{revtex4}%
\usepackage{amsfonts}
\usepackage{amsmath}
\usepackage{amssymb}
\usepackage{graphicx}%
\begin{document}
\title{{\Large Multi-twist optical M\"{o}bius strips}}
\author{{\large Isaac Freund}\medskip}

\affiliation{Physics Department, Bar-Ilan University, Ramat-Gan ISL52900, Israel\bigskip}

\begin{abstract}
Circularly polarized Gauss-Laguerre $GL_{0}^{0}$ and $GL_{0}^{1}$ laser beams
that cross at their waists at a small angle are shown to generate a
quasi-paraxial field that contains an axial line of circular polarization, a C
line, surrounded by polarization ellipses whose major and minor axes generate
multi-twist M\"{o}bius strips with twist numbers that increase with distance
from the C point. \ These M\"{o}bius strips are interpreted in terms of Berry's
phase for parallel transport of the ellipse axes around the C point.

\vspace*{-0.08in}\medskip\emph{OCIS codes:}\ 260.0260, 260.5430, 260.6042,
140.3460. \ \

\end{abstract}

\maketitle

\hspace*{-0.13in}We show here how to create quasi-paraxial, elliptically
polarized light fields in which the major and minor axes of the polarization
ellipses generate M\"{o}bius strips with radially increasing numbers of
half-twists; to our knowledge, these polarization structures are unique.

The polarization ellipses in elliptically polarized \emph{paraxial} fields lie
in parallel planes (here the $xy$-plane) oriented normal to the propagation
direction ($z$-axis). \ The generic point polarization singularity in any
plane of such a field is a C point, an isolated point of circular polarization
embedded in a field of polarization ellipses [$1-5$]. \ The azimuthal
orientation of a circle is undefined (singular), and as a result the
polarization ellipses surrounding a C point rotate about the point,
generically with winding number (net rotation angle/$2\pi$) $I_{C}=$ $\pm1/2$.
\ As the beam propagates in space the C point traces out a line, a C line
[$1-5$].

When the beam opens up and is no longer paraxial, the polarization ellipses
are no longer constrained to lie in parallel planes; C lines, however, remain.
In any plane pierced by a C line a C point appears, and the projections onto
the plane of the surrounding ellipses continue to rotate about the point with
winding number, or topological charge, $I_{C}=$ $\pm1/2$.

\begin{figure}
[pb]
\includegraphics[width=0.35\textwidth  ]%
{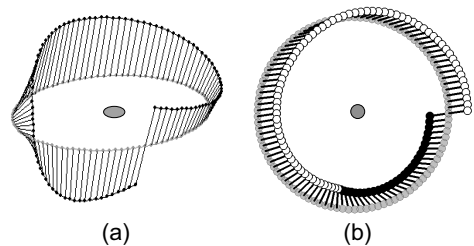}%
\caption{Computed M\"{o}bius strip surrounding a C point in a non-paraxial,
elliptically polarized optical field. \ (a) Three dimensional view of the
M\"{o}bius strip. Here the major axes of the ellipses on a circle centered on
the C point rotate through $180^{o}$ around the circle circumference to
generate a one-half-twist M\"{o}bius strip. \ Throughout, for clarity, only
half-axes are shown. \ (b) The strip in (a) seen from above. \ In this view
ellipse centers are shown by filled gray circles and semi-axes by straight
black lines; axis endpoints that lie above (below) the circle of ellipse
centers are shown by filled white (black) circles. \ The ellipse axes can be
seen to rotate counterclockwise about the circle circumference, forming a
half-turn segment of a left-handed circular screw.}%
\label{Fig1}%
\end{figure}

What is the full, three-dimensional (3D) arrangement of the ellipses that
surround a C line in a \emph{non}-paraxial field? \ The (possibly surprising)
answer is that the major and minor axes of these ellipses generate M\"{o}bius
strips [$6$]. \ The canonical M\"{o}bius strip has a single half-twist, and
can be either right-handed (RH) or left-handed (LH). \ Such strips have
recently been shown [$6$] to occur naturally in random fields. \ Illustrated
in Fig. \ref{Fig1} is an example of a left-handed M\"{o}bius strip. \ In
random fields left- and right-handed strips are equally probable.

M\"{o}bius strips with a larger, odd number of half-twists are possible, and
strips with three half-twists are also found in random fields [$6$]; in such
fields $\sim2/3$ of all M\"{o}bius strips have one half-twist, the remainder three
half-twists. \ Optical M\"{o}bius strips with more than three half-twists,
although not forbidden topologically, do not occur naturally.

Can such strips be generated? \ The answer is yes, and we show here how to
combine two laser beams to create a quasi-paraxial, elliptically polarized
beam containing a C line surrounded by polarization ellipses whose major and
minor axes generate M\"{o}bius strips with large numbers of half-twists. %

An elliptically polarized \emph{paraxial} laser beam containing an axial C
line can be generated by a coaxial, coherent superposition of a circularly
polarized RH (LH) Gauss-Laguerre $GL_{0}^{0}$ mode and a circularly polarized
LH (RH) $GL_{0}^{1}$ mode. \ The LH (RH)\ $GL_{0}^{1}$ mode contains a central
optical vortex (phase singularity) at which the amplitude vanishes [$7$], so
at that point the polarization in the combined beam is RH (LH) circular. \ At
other points in the beam RH and LH components combine to produce elliptical polarization.

Within the paraxial approximation, at the waists of the individual beams, and
therefore at the waist of the combined beam, there exist only the two
transverse field components $E_{x}$ and $E_{y}$. \ However, if the two beams
are made to intersect in the $xz$-plane at small angles $\pm\theta$ relative
to the $z$-axis, a third field component $E_{z}\sim E_{x}\sin\theta$ develops.
\ At its waist, the resulting field, described quantitatively below, has the
following unique property: \ Surrounding the central C point are circular
rings of radius $r$ on which the major and minor axes of the polarization
ellipses generate multi-twist M\"{o}bius strips that contain an odd number of
half-twists; these strips have the unique property that\emph{ the number of
half-twists increases with radial distance }$r$\emph{ from the C point}!

\begin{figure}
[h]
\includegraphics[width=0.25\textwidth]%
{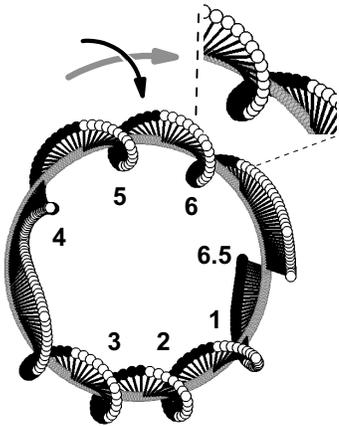}%
\caption{Computed multi-twist optical M\"{o}bius strip generated by the major
axes of the polarization ellipses on a circle surrounding a C point. \ An
observer walking (gray arrow) along this circle sees the ellipse axes rotate
in the clockwise direction (black arrow), generating a M\"{o}bius strip
containing $13$ half-twists that forms a circular, right-handed screw (helical
worm gear) containing $6.5$ turns (teeth).}%
\label{Fig2}%
\end{figure}

In Fig. \ref{Fig2} we show an example of such a M\"{o}bius strip. \ In
principle, $r$, and therefore the number of half-twists, increases without
limit; in practice, of course, the Gaussian envelope of the beam reduces the
intensity to immeasurably small levels far from the beam center.\vspace
*{-0.1in}%

\vspace*{-0.1in}

We turn now to a quantitative description of these
multi-twist optical M\"{o}bius strips. \ For the sake of definiteness, we take
the $GL_{0}^{0}$ mode to be to be RH and the vortex containing $GL_{0}^{1}$
mode to be left handed. \ We assume that both beams have the same wavelength
$\lambda$, and the same waist parameter $\mathfrak{w}_{0}\gg\lambda$, and that
they intersect maximally at their waists which are centered on the origin.
\ Although there are a number of different experimental approaches to
generating such beams, the one most likely to be used involves liquid crystal
modulators. \ These are usually addressed as a Cartesian grid of pixels, and
so in what follows we use Cartesian coordinates.

We define a Gaussian envelope function for a $GL_{0}^{l}$ beam, $l=0,1$,
by \vspace*{-0.2in}%

\begin{align}
G_{0}^{l}  &  =\left(  \mathfrak{w}_{0}/W_{0}^{l}\right)  B_{0}^{l}\exp\left(
-\left(  \rho_{0}^{l}\right)  ^{2}/\left(  W_{0}^{l}\right)  ^{2}\right)
\nonumber\\
&  \times\exp\left(  -ik\left(  \rho_{0}^{l}\right)  ^{2}/(2R_{0}^{l})\right)
\exp\left(  -ikZ_{0}^{l}\right)  , \label{Eq1}%
\end{align}
where $\left(  \rho_{0}^{l}\right)  ^{2}=\left(  X_{0}^{l}\right)  ^{2}%
+y^{2},R_{0}^{l}=Z_{0}^{l}+\mathfrak{Z}_{0}/Z_{0}^{l},k=2\pi/\lambda
,\mathfrak{Z}_{0}=k\mathfrak{w}_{0}^{2}/2,W_{0}^{l}=\mathfrak{w}_{0}%
\sqrt{1+\left(  Z_{0}^{l}/\mathfrak{Z}_{0}\right)  ^{2}},B_{0}^{l}=\left(
1+iZ_{0}^{l}/\mathfrak{Z}_{0}\right)  /\sqrt{1+\left(  Z_{0}^{l}%
/\mathfrak{Z}_{0}\right)  ^{2}}$. \ Writing $\theta_{0}^{l}$ for the angle
that the $GL_{0}^{l}$ beam makes with the $z$-axis, \vspace*{-0.2in}%

\begin{equation}
X_{0}^{l}=x\cos\theta_{0}^{l}+z\sin\theta_{0}^{l},\;Z_{0}^{l}=-x\sin\theta
_{0}^{l}+z\cos\theta_{0}^{l}. \label{Eq2b}%
\end{equation}

The field components of the combined beam $\mathbf{E=}E_{x}\mathbf{\hat{x}%
}+E_{y}\mathbf{\hat{y}}+E_{z}\mathbf{\hat{z}}$, with $\mathbf{\hat{x}%
},\mathbf{\hat{y}},\mathbf{\hat{z}}$ unit vectors along the corresponding
coordinate axes, are \vspace*{-0.2in}

\begin{align}
E_{x}  &  =\left(  E_{0}^{0}\right)  _{x}+\left(  E_{0}^{1}\right)
_{x},\;E_{y}=\left(  E_{0}^{0}\right)  _{y}+\left(  E_{0}^{1}\right)
_{y},\nonumber\\
E_{z}  &  =\left(  E_{0}^{0}\right)  _{x}\sin\theta_{0}^{0}+\left(  E_{0}%
^{1}\right)  _{x}\sin\theta_{0}^{1}, \label{Eq3c}%
\end{align}

\vspace*{-0.05in} where \vspace*{-0.2in}%

\begin{subequations}
\begin{align}
\left(  E_{0}^{0}\right)  _{x}  &  =G_{0}^{0},\;\left(  E_{0}^{0}\right)
_{y}=iG_{0}^{0},\label{Eq4b}\\
\left(  E_{0}^{1}\right)  _{x}  &  =\sqrt{2}G_{0}^{1}B_{0}^{1}\left(
X_{0}^{1}+i\sigma y\right)  /W_{0}^{1},\label{Eq4c}\\
\left(  E_{0}^{1}\right)  _{y}  &  =-i\sqrt{2}G_{0}^{1}B_{0}^{1}\left(
X_{0}^{1}+i\sigma y\right)  /W_{0}^{1}. \label{Eq4d}%
\end{align}
\end{subequations}

In Eqs. (\ref{Eq4c}) and (\ref{Eq4d}) $\sigma=+1$ ($\sigma=-1$) for a positive
(negative) vortex. \ For all figures presented here, $\mathfrak{w}%
_{0}=100\lambda$, $\theta_{0}^{0}=-\theta_{0}^{1}=5^{o}$, and $\sigma=+1$.

The major axis $\boldsymbol{\alpha}$ and minor axis $\boldsymbol{\beta}$ of
the ellipses surrounding the C point in Figs. \ref{Fig1} and \ref{Fig2} (and
Figs. \ref{Fig3} and \ref{Fig4} below), are obtained from Berry's formulas
[$5$]: $\boldsymbol{\alpha}=\operatorname{Re}\left(  \mathbf{E}^{\mathbf{\ast
}}\sqrt{\mathbf{E\cdot E}}\right)  $, $\boldsymbol{\beta}=\operatorname{Im}%
\left(  \mathbf{E}^{\mathbf{\ast}}\sqrt{\mathbf{E\cdot E}}\right)  $.

The exact $r$ dependence of the number of half-twists and handedness of the
M\"{o}bius strips requires an analyses that is too complicated to describe here.
\ Instead, insight into what happens can be obtained by considering the angle
$\alpha_{xy}=\arctan\left(  \alpha_{y},\alpha_{x}\right)  $ shown in Fig.
\ref{Fig3}, where $\alpha_{x}$ and $\alpha_{y}$ are the $x,y$-components of
$\boldsymbol{\alpha}$. \ Because $\boldsymbol{\alpha}$ is a line, not a
vector, it is plotted modulo $\pi$. \ On the small circle at the center of the
figure $\alpha_{xy}$ winds around the central C point, which in such plots
appears as a vortex, with winding number $I_{C}=-1/2$: $I_{C}=-1/2$ because
along a counterclockwise, by convention positive, circuit, $\alpha_{xy}$
decreases by $\pi$.%

\begin{figure}
[pth]
\includegraphics[width=0.2\textwidth]%
{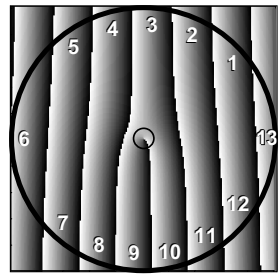}%
\caption{Rotation angle $\alpha_{xy}$, plotted $0$ to $\pi$ black to white, of
the projection of the major axes $\boldsymbol{\alpha}$ onto the $xy$-plane.}%
\label{Fig3}%
\end{figure}

\vspace*{-0.1in}The corresponding M\"{o}bius strip, shown in Fig. \ref{Fig1},
can be interpreted in terms of Berry's phase [$8$]: as one moves along the
circle surrounding the\ C point, $\boldsymbol{\alpha}$ undergoes parallel
transport, rotating through $\pi$ during one complete circuit; this leads to
winding number $I_{C}$ in the $xy$-projection, and the one-half-twist
M\"{o}bius strip in 3D.

In addition to the central C point, Fig. \ref{Fig3} contains a number of $\pi
$-fringes\ that are analogous to the $2\pi$-fringes in the forked-fringe
method for measuring vortices [$9,10$]. \ Each time $\boldsymbol{\alpha}$
passes through a $\pi$-fringe it rotates through $\pi$, so that the total
number of half-twists equals the total number of fringes traversed during a
circuit about the C point.\ The large circle in Fig. \ref{Fig3} that passes
through $13$ $\pi$-fringes corresponds to the M\"{o}bius strip in Fig.
\ref{Fig2}, which has $13$ half-twists. \ Similar considerations apply to axis
$\boldsymbol{\beta}$, and in Fig. \ref{Fig4} we show a M\"{o}bius strip
generated by this axis which has $29$ half-twists ($14.5$ full twists).%

\begin{figure}
[h]
\includegraphics[width=0.29\textwidth]%
{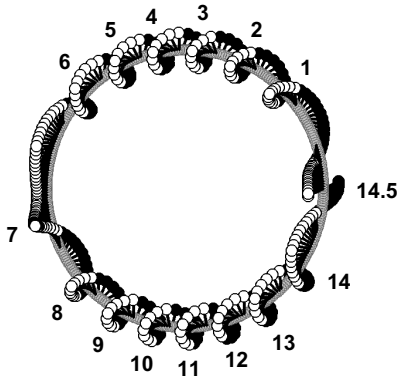}%
\caption{M\"{o}bius strip containing $29$ half-twists generated by minor axis
$\boldsymbol{\beta}$. \ Here radius $r=42\lambda$ is twice that of Fig. 2.}%
\label{Fig4}%
\end{figure}

\vspace*{-0.1in}The forgoing leads to a simple, heuristic expression for
twist number $\tau$, the number of \emph{full} twists, as a function of the radial
distance $r\ll\mathfrak{w}_{0}$ from the C point. \ For $\theta_{0}%
^{0}=-\theta_{0}^{1}=\theta$, far from the C point the fringe spacing along
the here horizontal $x$-axis, is $s=\lambda/\left(  2\sin\theta\right)  $, so
that the number of fringes traversed by a line of length $2r$, the circle
diameter, is $2r/s$. \ On a circle centered on the C point the circumference
passes through each fringe twice, and upon adding in the extra $\pi$-fringe
induced by the C point we obtain\vspace*{-0.20in}%

\begin{equation}
\tau_{calc}\simeq\text{int}(4r\sin\left(  \theta\right)  /\lambda
)+1/2.\label{Eq5}%
\end{equation}

\vspace*{-0.03in}In Fig. \ref{Fig1}, $r=2\lambda$, $\tau_{calc}=0.5$, in Fig.
\ref{Fig2}, $r=20\lambda$, $\tau_{calc}=6.5$, and in Fig. \ref{Fig4},
$r=42\lambda$, $\tau_{calc}=14.5$ $-$ in each case in agreement with Eq.
(\ref{Eq5}). \ We find Eq. (\ref{Eq5}) to be in general agreement with our
computer simulations for $r<\mathfrak{w}_{0}/2$; for larger $r$, $s$ begins to
decrease significantly with increasing $r$ due to wavefront curvature and the
Gouy phase shift, and Eq. (\ref{Eq5}) underestimates the twist number.

To be reported on elsewhere are the twist number and handedness of the
M\"{o}bius strips as a function of $r$ and other system parameters, the
$n$-foil knots generated by the axis endpoints of $n$-twist M\"{o}bius strips,
the strips (twisted ribbons) with an even number of half-twists that appear on
circles that do not enclose a C point, the multi-twist optical M\"{o}bius
strips that are generated by other combinations of laser modes, and by other
beam configurations, and the $z$ dependence of these structures.

In summary, we have shown how to combine two circularly polarized laser beams,
one of which contains a vortex, to create a quasi-paraxial field containing a
central C point\ surrounded by polarization ellipses whose major and minor
axes generate multi-twist optical M\"{o}bius strips with twist numbers that
increase radially from the C point. \ These M\"{o}bius strips are structurally
stable, changing unimportantly when, for example, $3\%$ noise is added to the
simulation. \ Coherent nanoprobe techniques [$11-18$] capable of determining
the field structure on subwavelength scales should permit experimental
measurements of these highly unusual objects. \ As for possible applications,
we note, inter alia, that the M\"{o}bius strips could be embedded in
polarization sensitive photoresists to create devices with unique optical
properties.\vspace*{0.08in}

\hspace{-0.15in}email address: freund@mail.biu.ac.il (I. Freund).

\vspace*{0.12in}

\hspace{-0.15in}{\large \textbf{References}}

\vspace*{0.07in}

\hspace{-0.15in}[1] J. F. Nye, \emph{Natural Focusing and Fine Structure of
Light} (IOP Publ., Bristol, 1999).

\hspace{-0.15in}[2] J. F. Nye, Proc. Roy. Soc. Lond. A \textbf{389}, 279$-$290 (1983).

\hspace{-0.15in}[3] M. V. Berry and M. R. Dennis, Proc. Roy. Soc. Lond. A
\textbf{457}, 141$-$155 (2001).

\hspace{-0.15in}[4] M. R. Dennis,\ Opt. Commun. \textbf{213}, 201$-$221 (2002).

\hspace{-0.15in}[5] M. V. Berry,\ J. Opt. A \textbf{6}, 675$-$678 (2004).

\hspace{-0.15in}[6] I. Freund, \textquotedblleft Optical M\"{o}bius strips in
three-dimensional ellipse fields: I. Lines of circular
polarization,\textquotedblright\ Opt. Commun. in press,
doi:10.1016/j.optcom.2009.09.042 (2009).

\hspace{-0.15in}[7] M. S. Soskin and M. V. Vasnetsov, \textquotedblleft Linear
theory of optical vortices,\textquotedblright\ in \emph{Optical Vortices}, M.
Vasnetsov and K. Staliunas Eds., Horizons in World Physics \textbf{228,}
1$-$35 (Nova Science Publs., Commack, New York, 1999).

\hspace{-0.15in}[8] An excellent review of Berry's phase in optics in given in
E. J. Galvez, \textquotedblleft Applications of geometric phase in
optics,\textquotedblright\ available from Wikipedia (Berry's phase), or http://departments.colgate.edu/physics/faculty/galvez.htm.

\hspace{-0.15in}[9] N. B. Baranova, B. Ya Zel'dovich, A. V. Mamaev, N.
Pilipetskii, and V. V. Shkukov, JETP Lett. \textbf{33}, 195$-$199 (1981).

\hspace{-0.15in}[10] I. V. Basistiy, M. S. Soskin, and M. V. Vasnetsov, Opt.
Commun. \textbf{119}, 604$-$612 (1995).

\hspace{-0.15in}[11] R. Dandliker, I. Marki, M. Salt, and A. Nesci,\ J. Optics
A \textbf{6}, S189$-$S196 (2004).

\hspace{-0.15in}[12] P. Tortora, R. Dandliker, W. Nakagawa, and L. Vaccaro,
\ Opt. Commun. \textbf{259}, 876$-$882 (2006).

\hspace{-0.15in}[13] C. Rockstuhl, I. Marki, T. Scharf, M. Salt, H. P. Herzig,
and R. Dandliker, \ Current Nanoscience \textbf{2}, 337$-$350 (2006).

\hspace{-0.15in}[14] P. Tortora, E. Descrovi, L. Aeschimann, L. Vaccaro, H. P.
Herzig, and R. Dandliker, Ultramicroscopy \textbf{107}, 158$-$165 (2007).

\hspace{-0.15in}[15] K. G. Lee, H. W. Kihm, J. E. Kihm, W. J. Choi, H. Kim, C.
Ropers, D. J. Park, Y. C. Yoon, S. B. Choi, H. Woo, J. Kim, B. Lee, Q. H.
Park, C. Lienau C, and D. S. Kim,\ Nature Photonics \textbf{1}, 53$-$56 (2007).

\hspace{-0.15in}[16] Z. H. Kim and S. R. Leone,\ Opt. Express \textbf{16},
1733$-$1741 (2008).

\hspace{-0.15in}[17] R. J. Engelen, D. Mori, T. Baba, and L. Kuipers, Phys.
Rev. Lett. \textbf{102}, 023902 (2009);Erratum: ibid. 049904 (2009).

\hspace{-0.15in}[18] M. Burresi, R. J. Engelen, A. Opheij, D. van Oosten, D.
Mori, T. Baba, and L. Kuipers, Phys. Rev. Lett. \textbf{102}, 033902 (2009).

\newpage

\hspace{-0.15in}{\large \textbf{References with titles}}

\bigskip

\hspace{-0.15in}[1] J. F. Nye, \emph{Natural Focusing and Fine Structure of
Light} (IOP Publ., Bristol, 1999).

\hspace{-0.15in}[2] J. F. Nye, \textquotedblleft Lines of circular
polarization in electromagnetic wave fields,\textquotedblright\ Proc. Roy.
Soc. Lond. A \textbf{389}, 279$-$290 (1983).

\hspace{-0.15in}[3] M. V. Berry and M. R. Dennis, \textquotedblleft
Polarization singularities in isotropic random vector waves,\textquotedblright%
\ Proc. Roy. Soc. Lond. A \textbf{457}, 141$-$155 (2001).

\hspace{-0.15in}[4] M. R. Dennis, \textquotedblleft Polarization singularities
in paraxial vector fields: morphology and statistics,\textquotedblright\ Opt.
Commun. \textbf{213}, 201$-$221 (2002).

\hspace{-0.15in}[5] M. V. Berry, \textquotedblleft Index formulae for singular
lines of polarization,\textquotedblright\ J. Opt. A \textbf{6}, 675$-$678 (2004).

\hspace{-0.15in}[6] I. Freund, \textquotedblleft Optical M\"{o}bius strips in
three-dimensional ellipse fields: I. Lines of circular
polarization,\textquotedblright\ Opt. Commun. in press,
doi:10.1016/j.optcom.2009.09.042 (2009).

\hspace{-0.15in}[7] M. S. Soskin and M. V. Vasnetsov, \textquotedblleft Linear
theory of optical vortices,\textquotedblright\ in \emph{Optical Vortices}, M.
Vasnetsov and K. Staliunas Eds., Horizons in World Physics \textbf{228,}
1$-$35 (Nova Science Publs., Commack, New York, 1999).

\hspace{-0.15in}[8] An excellent review of Berry's phase in optics in given in
E. J. Galvez, \textquotedblleft Applications of geometric phase in
optics,\textquotedblright\ available from Wikipedia (Berry's phase), or http://departments.colgate.edu/physics/faculty/galvez.htm.

\hspace{-0.15in}[9] N. B. Baranova, B. Ya Zel'dovich, A. V. Mamaev, N.
Pilipetskii, and V. V. Shkukov, \textquotedblleft Dislocations of the
wave-front of a speckle-inhomogeneous field (theory and
experiment),\textquotedblright\ JETP Lett. \textbf{33}, 195$-$199 (1981).

\hspace{-0.15in}[10] I. V. Basistiy, M. S. Soskin, and M. V. Vasnetsov,
\textquotedblleft Optical wavefront dislocations and their
properties,\textquotedblright\ Opt. Commun. \textbf{119}, 604$-$612 (1995).

\hspace{-0.15in}[11] R. Dandliker, I. Marki, M. Salt, and A. Nesci,
\textquotedblleft Measuring optical phase singularities at subwavelength
resolution,\textquotedblright\ J. Optics A \textbf{6}, S189$-$S196 (2004).

\hspace{-0.15in}[12] P. Tortora, R. Dandliker, W. Nakagawa, and L. Vaccaro,
\textquotedblleft Detection of non-paraxial optical fields by optical fiber
tip probes,\textquotedblright\ Opt. Commun. \textbf{259}, 876$-$882 (2006).

\hspace{-0.15in}[13] C. Rockstuhl, I. Marki, T. Scharf, M. Salt, H. P. Herzig,
and R. Dandliker, \textquotedblleft High resolution interference microscopy: A
tool for probing optical waves in the far-field on a nanometric length
scale,\textquotedblright\ Current Nanoscience \textbf{2}, 337$-$350 (2006).

\hspace{-0.15in}[14] P. Tortora, E. Descrovi, L. Aeschimann, L. Vaccaro, H. P.
Herzig, and R. Dandliker, \textquotedblleft Selective coupling of HE11 and
TM01 modes into microfabricated fully metal-coated quartz
probes,\textquotedblright\ Ultramicroscopy \textbf{107}, 158$-$165 (2007).

\hspace{-0.15in}[15] K. G. Lee, H. W. Kihm, J. E. Kihm, W. J. Choi, H. Kim, C.
Ropers, D. J. Park, Y. C. Yoon, S. B. Choi, H. Woo, J. Kim, B. Lee, Q. H.
Park, C. Lienau C, and D. S. Kim, \textquotedblleft Vector field microscopic
imaging of light,\textquotedblright\ Nature Photonics \textbf{1}, 53$-$56 (2007).

\hspace{-0.15in}[16] Z. H. Kim and S. R. Leone, \textquotedblleft
Polarization-selective mapping of near-field intensity and phase around gold
nanoparticles using apertureless near-field microscopy,\textquotedblright%
\ Opt. Express \textbf{16}, 1733$-$1741 (2008).

\hspace{-0.15in}[17] R. J. Engelen, D. Mori, T. Baba, and L. Kuipers,
\textquotedblleft Subwavelength Structure of the Evanescent Field of an
Optical Bloch Wave,\textquotedblright\ Phys. Rev. Lett. \textbf{102}, 023902
(2009); Erratum: ibid. 049904 (2009).

\hspace{-0.15in}[18] M. Burresi, R. J. Engelen, A. Opheij, D. van Oosten, D.
Mori, T. Baba, and L. Kuipers, \textquotedblleft Observation of Polarization
Singularities at the Nanoscale,\textquotedblright\ Phys. Rev. Lett.
\textbf{102}, 033902 (2009).

\end{document}